# Anomalous change in leakage and displacement currents after electrical poling on lead-free ferroelectric ceramics

Hitesh Borkar[1,2], M Tomar[3], Vinay Gupta[4], J. F. Scott[5], Ashok Kumar[1,2],*

[1]CSIR-National Physical Laboratory, Dr. K. S. Krishnan Marg, New Delhi 110012, India
[2]Academy of Scientific and Innovative Research (AcSIR), CSIR-National Physical Laboratory (CSIR-NPL) Campus, Dr. K. S. Krishnan Road, New Delhi 110012, India
[3]Department of Physics, Miranda House, University of Delhi, Delhi 110007, India
[4]Department of Physics and Astrophysics, University of Delhi, Delhi 110007, India
[5]Department of Chemistry and Department of Physics, University of St. Andrews, St. Andrews KY16 ST, United Kingdom

We report the polarization, displacement current and leakage current behavior of a trivalent nonpolar cation ($Al^{3+}$) substituted lead free ferroelectric $0.92(Na_{0.50}Bi_{0.50-x}Al_x)TiO_3$-$0.08BaTiO_3$ (NBAT–BT) (for x = 0, 0.05, 0.07, and 0.10) electroceramics with tetragonal phase and *P4mm* space group symmetry. Nearly three orders of magnitude decrease in leakage current were observed under electrical poling, which significantly improves microstructure, polarization, and displacement current. Effective poling neutralizes the domain pinning, traps charges at grain boundaries and fills oxygen vacancies with free charge carriers in matrix, thus saturated macroscopic polarization in contrast to that in upoled samples. E-poling changes "bananas" type polarization loops to real ferroelectric loops.

*Corresponding Author: Dr. Ashok Kumar: (Email: ashok553@nplindia.org)

Many reports on effect of electrical (E) poling on microstructures and functional properties of perovskite ($ABO_3$) electro-ceramics have been demonstrated in context of domain dynamics, domain switching from one to another crystal structure, charge migration, crystal densification, and grain boundaries reduction.[1,2,3] Over the last five decades effect of electrical poling on dielectric properties of $(1-x)(Na_{0.50}Bi_{0.50})TiO_3-xBaTiO_3$ (NBT-BT) have been studied in detail, most interesting observations are as follows: stabilization of long-range order and reduction of tilt disorder through compositional tunings, reduction of structural inhomogeneities, conversion of non-ergodic relaxor phase to a long-range polar phase, transformation of stripe-like unpoled nano-domains to lamellar tetragonal domains, switching of an antiferroelectric nonpolar to a ferroelectric polar state, improved piezo and ferroelectric properties under combined electrical and mechanical poling.[4,5,6,7,8,9,10] However the detailed microscopic conduction mechanisms before and after poling, particularly for the dia-, tri- and tetra- valent cations substituted systems remain missing so far.

Ferroelectric polycrystalline electroceramics possess permanent polarization below the Curie temperature ($T_c$) and its magnitude depends on the dipoles direction. Successive application of external electrical (E) fields produces local stress around grains and grain boundaries which leads to orientation of electrical dipoles in the direction of applied field. Ferroelectrics under constant E-field (poling) diminish local inhomogeneities in charged particles present.[11] The transport mechanisms in ferroelectrics have been studied in context of moment of dipole moments, charge transfer and trapping of charge carriers in domains and domain walls.[12,13,14,15] Particularly in polycrystalline ceramics, dipole-dipole interaction and dipole moments are complex in nature due to random orientation of mobile charge carriers in the grains, grain boundaries and across grain-grain boundary interfaces. Local inhomogeneities and movement of free charge carriers in the matrix make difficult to measure the actual displacement current. These electro-ceramics need strong E-fields to switch the polarization; in as grown samples, space charge limited current dominates in the electrical transport properties.

Several poling procedures such as high temperature poling, electromechanical poling, poling below coercive field, and short- and long- time poling, have been employed to orient the polarization in direction of applied electrical field, which significantly modifies the dipole orientation, domains restructuring, movement and trapping of charge carriers, and confinement of space charge in the grain boundaries.[4-12] The application of E-fields for a minimum exposure time is the most accepted procedure to get stable polarization. The best way to achieve the stable polarization in minimum time is heating of polar electroceramics to near $T_c$ and poled it for few hours (h) below the coercive field; in this situation the coercive field is very low compared to the normal values and therefore easy to pole.[16,17] Reports are also present in literature where moderate E-poling has been carried out on the paraelectric phase of the polar system to orient the dipoles in particular direction; in these conditions dipoles naturally align along the direction of the E-field during the cooling process in ferroelectric regions.[17-18] For electromechanical poling, mechanical stress and E-fields favor ferroelastic and ferroelectric switching, respectively, which in turn effectively poles the systems at low applied E-field with minimum built-in off-set polarization.[19] E-poling for long times can create a built-in bias E-field and off-set polarization in the samples that is mainly due to trapping of mobile charge carriers across the grains and grains-grain boundary interfaces. It would be hard to distinguish between the built-in polarization and the true polarization. In general researchers avoid long time E-poling in the samples for piezoelectric sensors and actuators applications.[20,21,22,23,24] However it is also true that built-in hysteresis loops are widely used to decrease hysteresis and self heating in hard piezoelectrics.[25]

T. Takenaka group has carried out extensive studies on polarization, phase transition, dielectric and electrical properties of (1-x)NBT-xBiAlO$_3$ solid solution with pre intended Al substitution at B-site.[26,27] The primary goal of this letter is to check the effect of E-poling on acceptor doped (oxygen vacancies) hard NBAT-BT electroceramic. We have shown that how E-poling significantly reduces leakage current and improves the displacement current. The poling was extremely effective for the Al-substituted NBT-BT systems compared to parent NBT-BT system.

Bulk polycrystalline (Na$_{0.46}$Bi$_{0.46-x}$Al$_x$Ba$_{0.08}$)TiO$_3$ (NBAT-BT) (x=0, 0.05, 0.07 and 0.10) electroceramics were prepared by a conventional ceramic processing technique. In the following the sample compositions of NBAT-BT for x=0 (S1), 0.05 (S2), 0.07(S3), 0.10 (S4) are abbreviated as S1, S2, S3, and S4, respectively. The high amount of Al substitution at A-site was chosen to check the considerable effect of possible nonpolar cation substitution on polarization, coercive field and leakage current which can be easily detected by SEM-EDX data within instrumental errors. It would be very hard to realize the replacement of small amount (< 4%) of Bi by Al substitution using the SEM-EDX. In supplemental material we provide the detail compositional variation of NBAT-BT systems with increase in Al concentrations (Table S1).[28] Precursor oxides Na$_2$CO$_3$, BaCO$_3$, Bi$_2$O$_3$, TiO$_2$ and Al$_2$O$_3$ from sigma aldrich (purity ~ 99.9%) were used to synthesize NBAT-BT systems. The mixed precursors of desired chemical formula were mechanically milled in a mortar and pestle with IPA (isopropyl alcohol) for 2 h for homogeneous mixing. Mixed powders were dried and calcined at 1000 °C for 4 h. Subsequently calcined powders were again mixed with a binder (10 wt% polyvinyl alcohol solution) in order to prepare circular disc shaped pellets for electrical and microstructural analysis. The binder-mix granulated powders were shaped into a circular discs of diameter 13 mm and thickness 1-1.5 mm under uniaxial pressure of 5–6 tons per square inch. Finally these pellets were sintered in air at 1200 °C for 8 h to achieve near the theoretical density. The bulk density of S1, S2, S3, and S4 sintered ceramics were found around 91.6 %, 90.5%, and 92.4%, and 93.8%, respectively, within the limit of experimental errors (+/- 2%).[7,23]

Phase purity and crystal structure were examined on sintered electroceramics by an XRD (Bruker AXS D8 Advance X-ray diffractometer) using the Cu-kα (kα=1.54059A°) monochromatic radiation, in the 2θ range between 20° to 80°. To measure ferroelectric and leakage properties, samples were carefully polished by mechanical pressure technique to make thin slabs of 0.5 mm and painted with silver paste on both surfaces. Silver painted pellets were dried at 250 °C for 2 h in order to get metal electrodes. Polarization versus electric field (P-E) hysteresis loops and leakage current were measured using Radiant Ferroelectric Tester. These pellets were poled at a high dc E-field of 40 kV/cm for 10 h to perform the electrical characterization. The tolerance factors were calculated for all electroceramics using Goldschmidt tolerance factor $t = \frac{\bar{r}_A + r_o}{\sqrt{2}(\bar{r}_B + r_0)}$ where $\bar{r}_A = \sum_{i=1}^{m} x_{Ai} r_{Ai}$  $\bar{r}_B = \sum_{i=1}^{m} x_{Bi} r_{Bi}$ , $x_{Ai}$ and $x_{Bi}$ are proportion of A-site and B-site elements, $r_{Ai}$ and $r_{Bi}$ are ionic radii of A-site and B-site elements, respectively. The t-values of NBAT-BT samples were found in the range of 0.95 to 0.92 and the value decreases with increasing Al concentration. These results indicate the formation of stable perovskite structure with Al substitution.[27]

Figure 1 shows the X-ray diffraction patterns of unpoled sintered pellets that exhibit pure perovskite structure with a small trace of secondary pyrocholre phases (< 0.1%). All three Al-substituted NBAT-BT systems showed the tetragonal crystal structure with noticeable shifts in Bragg angle towards higher angles; however, NBT-BT displayed a mixed monoclinic and tetragonal phase. All the peaks were indexed on the basis of tetragonal *P4mm* crystal structure.

Insets of Fig. 1 show decrease in the intensity of (110) and (002) planes and increase in the intensity of symmetric planes (101) and (200), respectively with increase in Al substitution. These crystal planes also show a shift towards higher angle side with increasing Al content which indicates that there is decrease in unit cell volume and lattice parameters. These results suggest that Al substituted NBAT-BT systems tend towards more pseudo-cubic crystal structure, owing to the relatively small ionic radius of $Al^{3+}$ compared to $Bi^{3+}$ ions. The amount of secondary phases were calculated using the integrated area occupied by the main XRD peak (110)/(101) and the secondary phases in XRD data. The percentage of secondary phases was calculated using the given formula; % secondary phase $= \frac{A_{SP}}{(A_{PP} + A_{SP})} X 100$, where $A_{SP}$ and $A_{PP}$ represent the integrated area under secondary phase and perovskite phase, respectively. The XRD peak near to 28.06 degree Braggs angle represents the secondary phase of $Bi_{48}Al_2O_{75}$ alloys (* JCPDS file No: 420199) and it was found nearly 0.1%, 0.3% and 0.1% in S2, S3 and S4 samples, respectively, however the XRD peaks near to 29.4 and 30.0 degree Braggs angle in the sample S4 provide nearly 0.5% secondary phase of possible $Na_xTi_yO_z$ alloys (# JCPDS file No: 370345 and 331295).[29] We believe that small amount of secondary pyrochlore phases may act as the trap center for charge carriers and affect the current conduction behavior and domain switching. The E-poling of Al-substituted NBT-BT samples significantly improved the microstructure, reduced the voids, defect densities, pining, and compaction among grains as compared with virgin samples. In supplemental material (Table S1) we show the systematic increase in wt% of Al and decrease in Bi as function of increasing Al substitution.[28]

Polarization-electric field (P-E) hysteresis loops and displacement currents D(t) were measured on unpoled and poled electro-ceramics at 60 kV/cm applied E-field with fixed probe frequency 10 Hz, as depicted in Fig. 2 (a-d) and Fig. 3 (a-d), respectively. Comparative values of remanent polarization ($P_r$) and coercive field ($E_c$) are listed in supplemental material (Table S2).[28] The Al substitutions at Bi site provides a noticeable increase in the coercive field of S2, S3, S4 samples, which implies hard (long-range) ferroelectrics compared to the basic cluster matrix. P-E loops for Al substituted unpoled NBT-BT ceramics show "ferroelectric banana type loops" with negligible displacement currents that may be entirely due to large leakage current, in contrast to real displacement current.[30] Unpoled S2, S3 and S4 samples exhibit primarily resistive type of current switching; however, after poling these samples show sharp peaks in displacement currents. The most interesting effect of E-poling was the improvement and evolution of "banana-type" unsaturated leakage current "loops" to well saturated real ferroelectric P-E loops. The question arises then: What happened microscopically to the microstructure of unploed ceramics so that the resistive current changes to displacement current? Basically, during the E-poling randomly oriented domains align in the direction of applied E-field, yielding larger net polarization and trapping of mobile free charge by the oxygen vacancies; this makes more stable octahedra with minimum local inhomogenities. It is also possible that Al cations act as possible defect centers which significantly reduce de-trapping of free charge carriers after E-poling. In contrast to S2, S3, S4 samples, pure NBT-BT shows well-saturated P-E loops before and after E-poling and hardly displays any change in the leakage current behavior with E-poling (Fig3(a)).[31] These results indicate that deficiencies of charge carriers at Al-based defect centers and associated oxygen vacancies are overcome by the free charge carriers during the E-poling. These free charge carriers get enough energy from external E-field to migrate from one place to another and compensate the deficiencies, making the octahedra more perfect and enhancing net dipole moments in the direction of applied field. E-poling effectively reduces the domain wall pinning and compensates defect density, which may be key factors to get well-saturated ferroelectric hysteresis loops.

Displacement current ($j_D$) of unpoled and poled electroceramics are shown in Fig. 3 (a-d). It can be noted that E-poled S2, S3, and S4 electroceramics show dipole-induced abrupt switching in displacement current peaks that change their direction according to applied electric field polarity. Large leakage currents (Fig.4 (a-d)) and associated low electric breakdown field in Al-substituted unpoled NBT-BT restrict the switching of ferroelectric domains and displacement current; thus only resistive-type current loops are observed. The leakage current density is reduced by approximately a factor of 2 or 3 under constant E-field poling as shown in Fig.4 (b-d).

To further investigate the conduction mechanisms for bulk electroceramics three main mechanisms have been considered, conventional Schottky emission (SE)[32], bulk-limited Poole–Frenkel (PF)[33] and space-charge-limited current (SCLC) conduction mechanisms.[34] Charge transport at high E-field is usually determined by the interface limited. Current density versus electric field relationships of SE and PF were plotted with *ln J vs. $E^{1/2}$* and *ln J/E vs. $E^{1/2}$*, respectively. Depending on the type of charge transport, experimental data should satisfy straight line fits, and theoretically the magnitude of slopes should represent the optical dielectric constant of materials. The magnitude of slopes were obtained from the straight line fit of the experimental data over large E-field regions as shown in Fig. 5 (a,b). These values are in disagreement with the real optical dielectric constant of parent NBT-BT systems and ruled out the presence of SE and PF mechanisms.[35] The experimental data were plotted for SCLC mechanism on unpoled (Fig.5 (c)) and poled (Fig.5 (d)) systems as function of *log(j)-log(E)*, the values of the slopes were calculated from the linear fit and it was found in the range of 1.1 for unpoled S2, S3, and S4 samples; we emphasize in what follows that such a linear slope is actually compatible with the Simmons' modification of the basic Schottky Equation. The usual Schottky Equation assumes that the electron mean free path is long compared with the Schottky barrier width. But in oxides the reverse is true. Therefore a linear dependence of *j(V)* cannot discriminate by itself between Simmons-Schottky emission and space charge limited currents in the low-voltage regime. However, the current-voltage slope is found to be 1.5 for poled S2, S3, and S4 samples, and in both poled and unpoled samples of pure NBT-BT, the slopes of j(V) are approximately 2. This favors SCLC over Simmons-Schottky emission. Trap charge-assisted current conduction was observed with increase in applied E-field. In the case of discrete trap conduction in these systems, current density is expected to follow the equation.[36]

$$j_{SCLC} = \frac{9\mu_p \varepsilon_r \varepsilon_o \theta E^2}{8d} \quad \text{............................................................................} \quad (1)$$

where, E is bias field, $\varepsilon_r$ is the dielectric constant of ceramics, $\varepsilon_o$ is the dielectric constant of free space, $\mu_p$ is carrier mobility, $\theta$ is the ratio of total density of induced free carriers to trapped carriers. If the internal built-in field is dominated by space charge carriers (either from free or trapped carriers), than current density $j_L$ should follow power-law dependence on electric field (J α $E^n$). In Fig. 5(d) the values of the slopes obtained from fitted parameters of poled samples were found in the range of ~ 1.5-2 which indicates that current conduction process was more near to discrete trap carriers. Our findings suggest that E-poling can relocate the anion (oxygen) vacancies and defects trapped in grains and grain boundaries.

In conclusion, nonpolar addition of Al in NBT-BT matrix increased by x1000 the space charge supported leakage current in unpoled electroceramic samples. These unpoled samples show "banana type" leakage-artifact hysteresis with almost resistive-type current; however, after prolonged E-poling, conduction current was drastically suppressed, and the specimens display well saturated P-E loops with sharp capacitive displacement current peaks. The nonpolar (Al) substitution at the Bi-site also destroys the localized relaxor-like polar-regions and provides a normal ferroelectric with moderate high coercive field. These studies extend the earlier work on Pb-based oxide ferroelectrics to the lead-free NBT-BT system.

**Acknowledgement:**

AK acknowledges the CSIR-MIST (PSC-0111) project for their financial assistance. Hitesh Borkar would like to acknowledge the CSIR (JRF) to provide fellowship to carry out Ph. D program. Authors would like to thank Dr. V N Ojha (Head ALSIM) and Dr. Sanjay Yadav, for their constant encouragement.

**Figure captions**

**Fig1.** XRD patterns of NBAT-BT electroceramics having chemical compositions $(Na_{0.46}Bi_{0.46-x}Al_xBa_{0.08})TiO_3$ (NBAT-BT) for x=0 (S1), 0.05 (S2), 0.07 (S3), 0.1 (S4) respectively. Inset shows magnified XRD patterns in the vicinity of (101) and (200) diffraction peaks of ceramics near Bragg's angle 32° and 46°, respectively. Star marks (*) in S2, S3, and S4 represent the secondary pyrochlore phases.

**Fig2.** Ferroelectric polarization measured at 10 Hz on NBAT-BT samples for unpoled and poled electroceramics (a) x=0 (b) x=0.05 (c) x=0.07, (d) x=0.1

**Fig3.** Displacement current derived form ferroelectric polarization for NBAT-BT for unpoled and poled electroceramics (a) x=0 (b) x=0.05 (c) x=0.07, (d) x=0.1

**Fig4.** Leakage current density (J) as function of electric field (E) of NBAT-BT systems for unpoled and poled electroceramics (a) x=0, (b) x=0.05, (c) x=0.07 and (d) x=0.1

**Fig5.** Conduction mechanisms characterization of NBAT-BT samples in the light of SE, PF and SCLC process for electrically poled electroceramics and fitted with (a) ln (J) vs $E^{1/2}$ plots, (b) ln (J/E) vs $E^{1/2}$ plots, and (c) log (J) vs log (E) plots for SCLC.

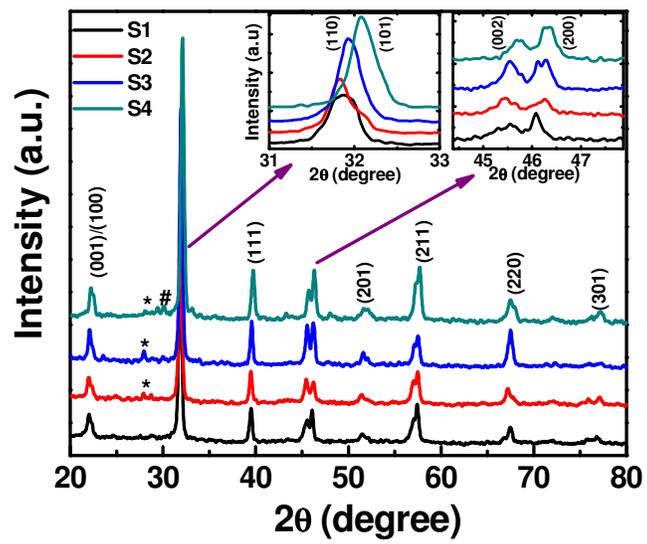

Fig.1

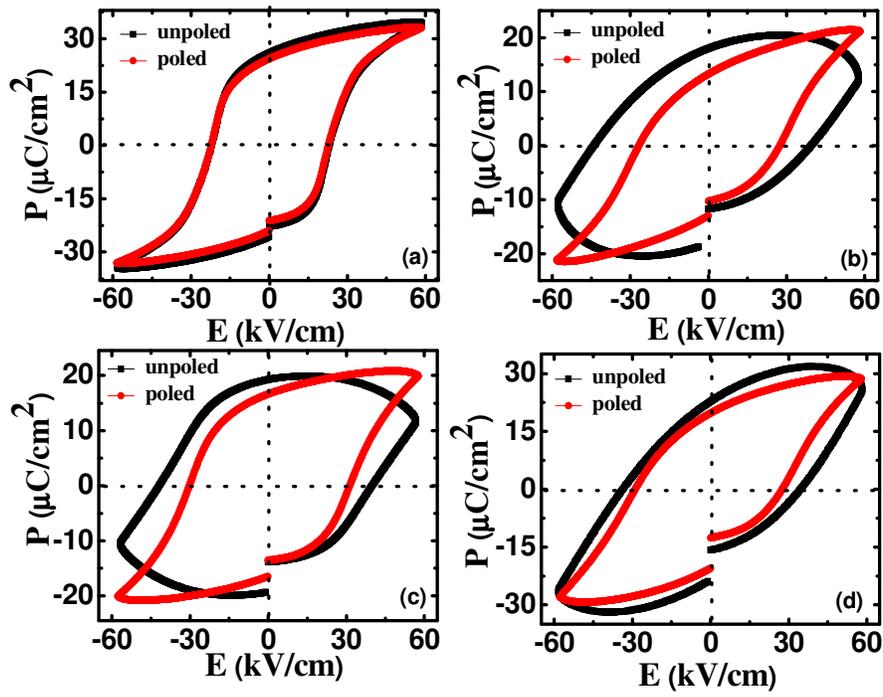

**Fig.2**

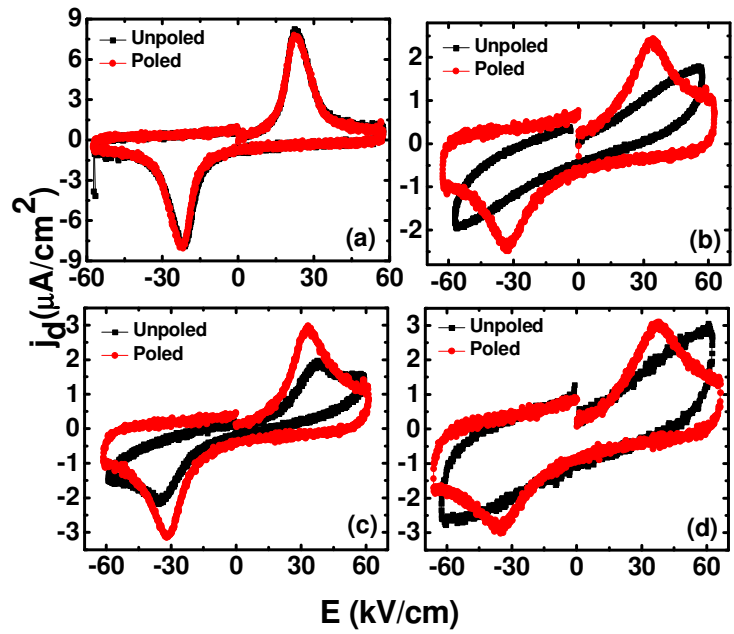

**Fig.3**

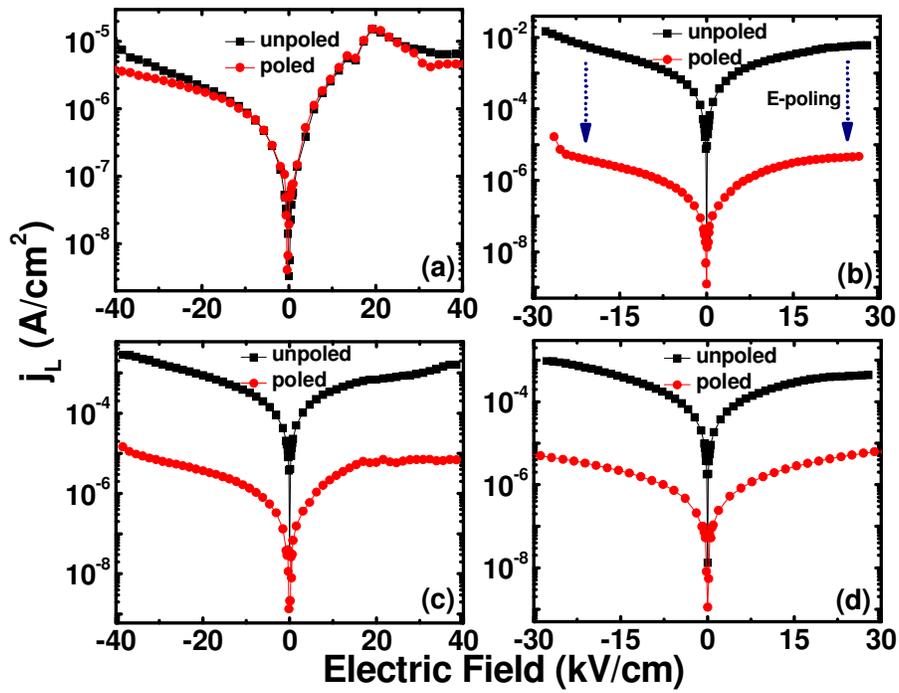

**Fig.4**

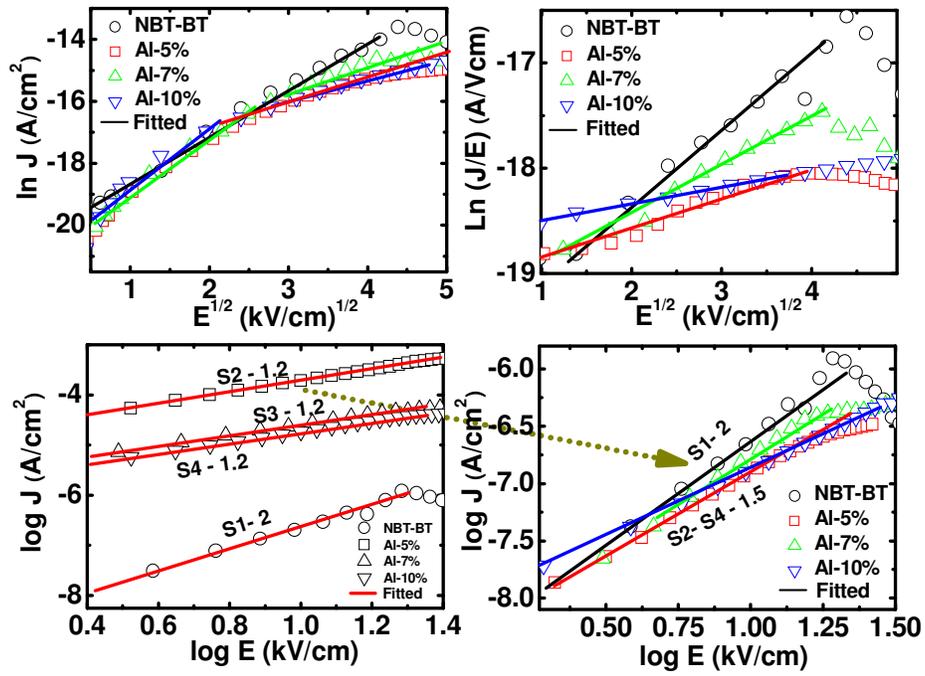

**Fig.5**